%% file: incgc.tex
\documentclass{tlp}
\usepackage{float}
\usepackage{epsfig}
\usepackage{subfigure}
\usepackage{xspace}
\usepackage{booktabs}

\newcommand{\gc}{garbage collection\xspace}

\newcommand{\earlyreset}{\textit{early reset}\xspace}
\newcommand{\Earlyreset}{\textit{Early reset}\xspace}
\newcommand{\partialer}{\textit{partial early reset}\xspace}
\newcommand{\tidytrail}{\textit{tidy trail}\xspace}
\newcommand{\Tidytrail}{\textit{Tidy trail}\xspace}

\newcommand{\remset}{remembered set\xspace}
\newcommand{\remsets}{remembered sets\xspace}

\newcommand{\fromspace}{\textit{from space}\xspace}
\newcommand{\tospace}{\textit{to space}\xspace}

\newcommand{\unsafe}{\textit{unsafe}\xspace}
\newcommand{\safe}{\textit{safe}\xspace}

\newcommand{\interblock}{\textit{inter-block}\xspace}
\newcommand{\intrablock}{\textit{intra-block}\xspace}

\newcommand{\fromblock}{\textit{from block}\xspace}
\newcommand{\toblock}{\textit{to block}\xspace}
\newcommand{\cublock}{\textit{current block}\xspace}

\newcommand{\sourceblock}{\textit{source block}\xspace}
\newcommand{\targetblock}{\textit{target block}\xspace}

\newcommand{\optgc}{\textbf{\textit{opt\_gc}}\xspace}
\newcommand{\incgc}{\textbf{\textit{inc\_gc}}\xspace}
\newcommand{\gengc}{\textbf{\textit{gen\_gc}}\xspace}

\newcommand{\oldheap}{\textbf{\textit{wam\_heap}}\xspace}
\newcommand{\newheap}{\textbf{\textit{bb\_heap}}\xspace}
\newcommand{\newheapwb}{\textbf{\textit{bb\_heap$\it{_{wb}}$}}\xspace}

\newcommand{\oldheaptab}{\textbf{\textit{wam}}\xspace}
\newcommand{\newheaptab}{\textbf{\textit{bb}}\xspace}
\newcommand{\newheapwbtab}{\textbf{\textit{bb$\it{_{wb}}$}}\xspace}

\newcommand{\boyer}{\textbf{boyergc}\xspace}
\newcommand{\browse}{\textbf{browsegc}\xspace}
\newcommand{\chess}{\textbf{chess}\xspace}
\newcommand{\dnamatch}{\textbf{dnamatchgc}\xspace}
\newcommand{\mqueens}{\textbf{mqueens}\xspace}
\newcommand{\serial}{\textbf{serialgc}\xspace}
\newcommand{\tak}{\textbf{takgc}\xspace}

\floatstyle{ruled}
\newfloat{Algorithm}{ht}{lop}

     {\begin{list}{}{
     \setlength{\itemsep}{0 pt}
     \setlength{\parsep}{0 pt}
     \setlength{\topsep}{2 pt} }}
     {\end{list}}

\newenvironment{myitemize}{%
  \begin{itemize}}{\end{itemize}}

\submitted{1 Februari 2005}
\revised{22 September 2005}
\accepted{22 December 2005}

\begin{document}


\title[Incremental Copying Garbage Collection for WAM-based Prolog systems]
{Incremental Copying Garbage Collection for WAM-based Prolog systems}

\author[Vandeginste et al.]
{RUBEN VANDEGINSTE, BART DEMOEN \\
  Dept.~of Computer Science, K.U.Leuven, Belgium\\
  \email{\{ruben,bmd\}@cs.kuleuven.ac.be}
}



\thispagestyle{empty}
\maketitle


\begin{abstract}
  The design and implementation of an incremental copying heap garbage
  collector for WAM-based Prolog systems is presented. Its heap layout
  consists of a number of equal-sized blocks. Other changes to the
  standard WAM allow these blocks to be garbage collected
  independently. The independent collection of heap blocks forms the
  basis of an incremental collecting algorithm which employs copying
  without marking (contrary to the more frequently used mark\&copy or
  mark\&slide algorithms in the context of Prolog). Compared to
  standard semi-space copying collectors, this approach to heap
  garbage collection lowers in many cases the memory usage and reduces
  pause times. The algorithm also allows for a wide variety of garbage
  collection policies including generational ones. The algorithm is
  implemented and evaluated in the context of hProlog.
\end{abstract}


\begin{keywords}
memory management of logic programming languages, incremental copying
garbage collection, WAM based Prolog implementation.
\end{keywords}


\section{Introduction}
\label{sec:introduction}


We will assume some familiarity with Prolog and its implementation.
The WAM \cite{DHWa83,AitK90} (Warren Abstract Machine) is a well-known
virtual machine with a specialised instruction set for the execution
of Prolog code. The WAM has proven to be a good basis for efficient
Prolog implementations, and many systems are based on it. We also
assume basic knowledge about garbage collection in general; a good
overview is given in \cite{wilson92uniprocessor,JonesLins}. Garbage
collection for Prolog is discussed in
\cite{SicstusGarbage@CACM-88,BekkersRidouxUngaro@IWMM-92,BevemyrLindgren@PLILP-94,DemoenNguyenVandeginsteICLP2002}.

The WAM defines three memory areas: a merged stack for environments
and choice points, a trail, and a heap. Different memory management
techniques are available to recover space in these areas. In this work
we focus on memory management of the heap. The basic WAM already
provides a mechanism to recover space allocated on the heap. Each time
backtracking occurs, all data allocated since the creation of the most
recent choice point can be deallocated. If we define a heap segment as
the space on the heap, delimited by the heap pointers in two consecutive
choice points, then upon backtracking, the WAM can recover all space on
the heap allocated for the most recent heap segment. This technique
for recovering heap space is called instant reclaiming.

In practice however, instant reclaiming alone is not
sufficient. Moreover, many Prolog programs are deterministic or the
backtracking is mainly shallow and in these cases instant reclaiming
is not effective. Therefore Prolog systems need garbage
collection for the heap. Early on, many systems used mark\&slide
garbage collection \cite{SicstusGarbage@CACM-88} because of the
following properties. Mark\&slide garbage collection preserves the cell
order and as a consequence also preserves heap segments (which is
important for instant reclaiming). Another important issue is memory
usage: besides mark and chain bits (2 extra bits per heap cell), no
extra space is needed for mark\&slide based garbage collection.

Recently copying garbage collection has become more popular.
\,Contrary to mark\&slide garbage collection, copying garbage
collection does not preserve the cell order, nor the heap segments and
consequently loses the ability to do instant reclaiming (see
\cite{DemoenEngelsTarau@SAC-96} for an exception).  A copying garbage
collector, however, has better complexity than a mark\&slide garbage
collector: mark\&slide collection is $O(n)$ with $n$ the
\textit{total} number of heap cells, while copying collection is
$O(m)$ with $m$ the number of \textit{surviving} heap cells.  Since
the fraction of heap cells that survive a collection is typically small
in Prolog systems, copying collectors in most cases outperform
mark\&slide collectors.  This can be observed in the benchmarks from
\cite{BevemyrLindgren@PLILP-94}, that compare the performance of
mark\&slide and copying garbage collectors in the context of a Prolog
system.

\begin{figure}[h]
  \begin{centering}
    \subfigure[Before copying]%
    {\epsfig{file=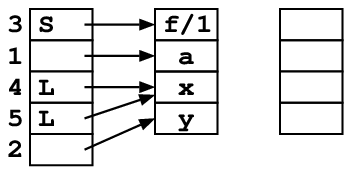,width=0.32\textwidth}%
      \label{optgcA}}
    \subfigure[After copying 1 \& 2]%
    {\epsfig{file=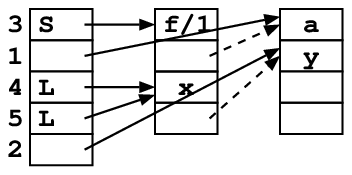,width=0.32\textwidth}%
      \label{optgcB}}
    \subfigure[After copying 3, 4 \& 5]%
    {\epsfig{file=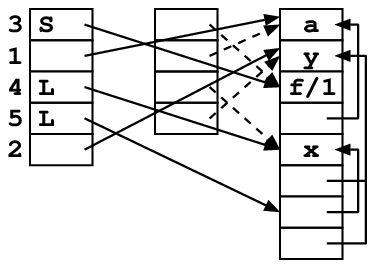,width=0.32\textwidth}%
      \label{optgcC}}
    \caption{Double and multiple copying}
    \label{optgc}
  \end{centering}
\end{figure}

Still, there is an issue with copying collectors in a Prolog
implementation based on the WAM: under certain conditions, some heap
cells are copied twice or even more during the collection and this can
cause the \tospace to overflow.  Therefore, these collectors are
considered \unsafe in the context of Prolog.  A detailed discussion of
this problem can be found in \cite{DemoenNguyenVandeginsteICLP2002}.
We illustrate this here with Figure~\ref{optgc}.  The figures show
from left to right: the root set, the \fromspace and the \tospace.
Dashed arrows indicate forwarding pointers.  The \texttt{S} stands for
\texttt{STRUCT} and the \texttt{L} for \texttt{LIST}.  In a semi-space
copying algorithm, heap cells referenced from the root set are copied
from the \fromspace to the \tospace and a forwarding pointer is left
in the \fromspace.  The order in which cells are forwarded is a priori
not known. In the figure, we indicated the order in which the root
cells are visited with the numbers on the left. The figures show that
if the argument ($a$) of a functor ($f/1$) is referenced independently
and copied to the \tospace (see Figure~\ref{optgcB}) before the
structure as a whole ($f(a)$), an extra reference (indirection to $a$)
is created in the \tospace (see Figure~\ref{optgcC}); this is called
\textit{double copying}.  Double copying also happens when one of the
cells of a list pair ($y$) is copied before the list pair ($[x|y]$) is
copied as a whole. But, as can be seen in Figure~\ref{optgcC}, it can
be even worse: when several \texttt{LIST} cells point to the same list
pair, extra copies of the list cells ($x$ and $y$) are created for
each additional \texttt{LIST} cell; this is called \textit{multiple
  copying}.  Multiple copying happens with \texttt{LIST} cells,
because lists do not have a \textit{header cell} which can be used to
store a forwarding pointer (for structures, the functor cell is used
to store a forwarding pointer; see Figure~\ref{optgcC}) and as such it
is not known during a collection whether (and whereto) a list pair has
already been forwarded as a whole.

\cite{BevemyrLindgren@PLILP-94} prevents double and multiple copying
by adding a marking phase before the copying phase.  This makes
copying collectors in the context of Prolog \safe.  Still, the marking
phase adds a considerable cost to the copying algorithm and recently,
work has been done to make copying without marking safer.
\cite{DemoenNguyenVandeginsteICLP2002} shows that double copying
occurs rarely, and that a small change to the copying algorithm
prevents multiple copying completely without marking (i.e., a list
cell is never copied more than twice).  In addition, an even simpler
change is presented, which prevents multiple copying in a very common
case.

\begin{figure}[h]
  \begin{centering}
    \subfigure[Before copying]%
    {\epsfig{file=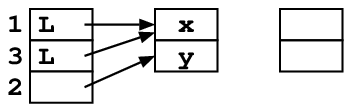,width=0.31\textwidth}%
      \label{optgcD}}
    \subfigure[After copying 1 \& 2]%
    {\epsfig{file=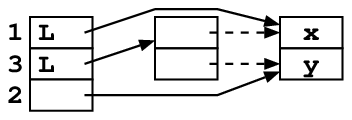,width=0.31\textwidth}%
      \label{optgcE}}
    \subfigure[After copying 3]%
    {\epsfig{file=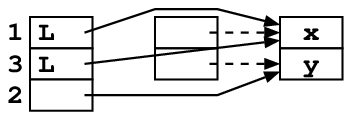,width=0.31\textwidth}%
      \label{optgcF}}
    \caption{Double and multiple copying}
    \label{optgc2}
  \end{centering}
\end{figure}

If, upon copying from a \texttt{LIST} cell, the two cells of the list
pair have already been forwarded to two consecutive locations, then
there is no need to create extra copies of them.  This is shown in
Figure~\ref{optgc2}.  The list pair ($[x|y]$) is first copied as a
whole (from \texttt{LIST} cell 1) and both list cells ($x$ and $y$)
are forwarded to two consecutive locations in the \tospace; see
Figure~\ref{optgcE}.  This means that upon subsequent copying from
another \texttt{LIST} cell (cell 3) referencing the same list pair,
this cell can be relocated to point to the forwarded list pair and no
extra cells are created; see Figure~\ref{optgcF}.  The copying
algorithm can be easily adapted to check whether the list cells have
been forwarded to consecutive locations upon copying from a
\texttt{LIST} pair.  The resulting algorithm is called
\textit{optimistic copying}.  Optimistic copying is, in practice,
successful in reducing the amount of duplication due to multiple
copying, while incurring no noticeable overhead. We use
\textit{optimistic copying} as the basis for our incremental
algorithm.

A simple copying collector can only perform \textit{major}
collections, collecting the full heap at once.  Since the Prolog
program is stopped during the collection cycle, this leads to big
pause times for applications with big heaps.  Some applications
however have timing constraints and require these pause times to be
small. Because of this, reducing pause times has always been an
important topic in garbage collection. In this work we reduce the
pause times by collecting only part of the heap during each collection
cycle. This is called incremental collection. An extra benefit of
incremental collection in the case of copying collection is that the
\tospace is smaller. Most copying collectors are semi-space copying
collectors and they need both a \fromspace for allocation and a
\tospace for collection. The \tospace must always be as big as the
\fromspace. This means however that only half of the allocated space
can be used as useful heap space. In the case of an incremental
collector, the \tospace only needs to be as big as the part to be
collected in the next collection cycle.
\\


In Section \ref{sec:heaplayout} we discuss how we modify the standard
heap layout of the WAM. We explain the modifications needed for
backtracking, trailing, variable binding and tidy trail in this new
layout. We also show how instant reclaiming can be done on parts of
the heap that have not been collected.  Next, in Section
\ref{sec:incremental} the implementation of the incremental garbage
collector is discussed. We introduce the notions of write barrier and
remembered set, which are both needed to guarantee correct incremental
collections. We also explain the interaction between remembered sets
and backtracking and the changes needed for correct backtracking.
Next, we discuss early reset and our approximation of it, \partialer.
Section \ref{sec:policies} contains a description of the implemented
\gc policies.  We evaluate the performance of both the incremental and
the generational collector in Section \ref{sec:experiments}. Some
benchmarks are presented to investigate the time performance as well
as the memory usage.  In Section \ref{sec:relatedwork} we discuss
related work.  Finally, we conclude with Section \ref{sec:conclusion}.
\\


The experimental evaluation in this paper was performed on a Pentium4
1.8Ghz (512Kb L2 cache) with 512Mb RAM. Timings are given in
milliseconds, space measurements in heap cells (4 bytes). The
incremental collectors have been implemented for hProlog 1.7. hProlog
is a successor of dProlog \cite{wamvariations} and is meant to become
a back-end of HAL \cite{HALoverview}. hProlog is based on the WAM, and
differs mainly in the fact that it has a separate choice point and
environment stack and that it always allocates free variables on the
heap (i.e. environment slots never contain an free variable). hProlog
1.7 uses by default \textit{optimistic copying}
\cite{DemoenNguyenVandeginsteICLP2002} for garbage collection. We will
refer to the systems with the incremental collector as \incgc and
\gengc, and to the original system with the \textit{optimistic}
collector as \optgc.  For the performance evaluation, \earlyreset was
enabled in \optgc and \partialer (as discussed in Section
\ref{subsec:partialearlyreset}) in \incgc and \gengc.

Our benchmark set consists of the following benchmarks: \chess,
\mqueens, \browse, \boyer, \dnamatch, \tak and \serial. \chess is a
slightly modified version of a program used in \cite{vitoriclp2001}.
This version was used before in
\cite{DemoenNguyenVandeginsteICLP2002}. \mqueens was also used in
\cite{DemoenNguyenVandeginsteICLP2002}. The other benchmarks are taken
from \cite{Li@PPDP-00}. They are classical benchmarks, but have extra
parameters to increase the size of the benchmark.  This makes them
more interesting for testing garbage collector performance.  We run
them with the following input: \browse (5000), \boyer (5), \dnamatch
(1000), \tak (28,16,8), \serial (1000000).

\begin{table}[h]
  \center
  {\small
    \input{tn_table_benchmarks}
  }
  \caption{Benchmark characteristics}\label{benchmarks}
\end{table}

Table \ref{benchmarks} contains information about the memory
requirements of the benchmarks.  For each benchmark, we measured the
amount of space it needs in the hProlog1.7 system to run without
triggering garbage collections or stack expansions. The table shows
the needed amount of space in the heap, the trail, the environment
stack and the choice point stack (in number of cells). Both \chess and
\mqueens cannot run without garbage collection: \chess explicitly
calls the garbage collector and \mqueens runs out of heap space
without \gc. From those numbers we can see that most benchmarks are
deterministic and have a small root set.

\section{Prolog execution with a modified heap layout}
\label{sec:heaplayout}

\subsection{\newheap: a block-based heap layout}
\label{subsec:heapblocks}

Our incremental garbage collector is based on a modified heap layout
which lends itself better to incremental collections. Instead of
having one (large) contiguous heap area, this new heap layout consists
of a number of (possibly small) heap blocks, each with a fixed number
of cells.  We will refer to this new heap layout as the \newheap, and
to the standard WAM heap layout as the \oldheap.

In \newheap, the \textit{logical} heap is an \textit{ordered} set of
heap blocks.  The blocks are kept chronologically ordered, i.e. by
creation time, independently of the address order. This is achieved by
linking the blocks in a doubly linked list; the heap is a chain of
blocks.  This is needed for backtracking and instant reclaiming (see
Section \ref{subsubsec:instantreclaiming}). In addition, each block
also contains a time stamp for fast comparison of block ages (see
Section \ref{subsubsec:trailing}).  Related to this is the need to
determine the block to which a cell belongs.  For this reason, blocks
are always allocated on an address which is a power of two: the block
to which a cell belongs, can easily be determined by masking the least
significant bits of the cell's address.

\begin{figure}[h]
\begin{centering}
\epsfig{file=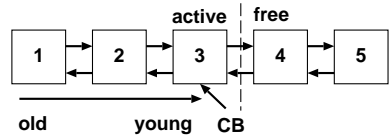,width=2in}
\caption{Doubly linked list of heap blocks}
\label{block_cfg}
\end{centering}
\end{figure}

The most recent block in the \newheap is used for allocation; we call
this the \cublock.  Without garbage collection, the heap is expanded
whenever the \cublock overflows: an extra heap block is allocated and
added to the heap.  Heap blocks that are part of the heap, are called
{\em active}. Heap blocks can become inactive because of instant
reclaiming or garbage collection. Inactive blocks are no longer part
of the heap: they are added to a free list and can be used for future
heap expansion.  In the implementation, the free list is appended to
the doubly linked list of heap blocks.  The chain of blocks then looks
like shown in Figure~\ref{block_cfg}.  The \texttt{CB} in the figure
stands for \cublock.  The \cublock (which \textit{always} contains the
top of the heap) functions as a separation between the active and the
inactive blocks.

The idea to divide the heap in separate blocks is not new: incremental
copying collectors as in \cite{hudson92incremental,beltway02} have a
very similar heap layout as the one we present here. One important
difference though is that \newheap keeps a strict order on the heap
blocks. In the context of WAM-based Prolog systems, this order is
important since the WAM uses the heap in a stack-like manner for
certain operations like backtracking and conditional trailing.
We know of two more LP-related systems that use a non-contiguous heap:
Quintus Prolog \cite{privateQuintus} employs a heap consisting of
equal sized heap blocks. Order between cells as required by the WAM
invariants is kept by moving the contents of the blocks (whenever
necessary) instead of by a time stamp, and the blocks are not used for
incremental garbage collection. 
The Oz implementation \cite{scheidhauerphd,mehlphd} uses a segmented
heap consisting of several chunks of memory of varying sizes. In this
implementation, cell order is not important and the heap layout is not
used for incremental garbage collection.

\subsection{Heap blocks during execution}
\label{subsec:backtrack}

\subsubsection{Instant reclaiming}
\label{subsubsec:instantreclaiming}

Instant reclaiming allows to recover an unbounded amount of
memory at a constant cost. It is therefore important to preserve
instant reclaiming during normal program execution, i.e. as long as
garbage collection has not taken place. Instant reclaiming relies on
the fact that the order of the heap segments is preserved. The
\newheap keeps the heap data chronologically ordered (as long as no
garbage collection has occurred), and as such also keeps the heap
segments ordered. This means that instant reclaiming is possible
within a block as well as over block boundaries. Upon backtracking all
heap cells allocated after the creation of the most recent choice point
(all heap cells belonging to the topmost heap segment) can be
deallocated easily. If the topmost segment is spread out over more
than one block, then all blocks that contain cells belonging to that
segment exclusively can be freed. In the block where this segment
starts, part of the block belonging to the reclaimed segment can also
be reclaimed.

Instant reclaiming in this manner is still a constant cost operation.
The following actions, which all have a constant cost, are done upon
backtracking: (a) the heap pointer $H_{CP}$ is retrieved from the most
recent choice point, (b) the block which contains $H_{CP}$, is
determined from its address and finally, (c) both \cublock and the top
of heap pointer $H$ are set.  Note that while instant reclaiming
itself is a constant cost operation, it does incur an extra cost,
which is not constant, in the final system. This will be discussed in
Section~\ref{subsec:remsetsbacktrack}.

\subsubsection{Trailing}
\label{subsubsec:trailing}

Upon backtracking to a certain choice point, all bindings done since
the creation of that choice point need to be undone. Trailing is the
WAM mechanism which remembers these bindings. The WAM only records the
relevant bindings on the trail: the binding of variables older than
the current choice point; this is achieved by conditional trailing.

In the \oldheap conditional trailing is easily done by comparing
the address of the variable which is about to be bound and the heap
pointer in the topmost choice point, as shown in the following
pseudo-code.
\vspace{-1mm}
\begin{verbatim}
   if (CellPtr < BH) trail(CellPtr);
\end{verbatim}
\vspace{-1mm}
If the variable is older (smaller address) than the
pointer in the choice point (points to higher address), then the
binding should be trailed. This relies on the fact that the contiguous
\oldheap grows towards higher memory addresses.

In the \newheap, things get a little more complicated because one
cannot rely on the fact that a higher memory address corresponds to a
more recent creation time.  As long as a block has not been garbage
collected, higher memory addresses still correspond to newer cells for
cells within that block. For cells in different blocks, one needs to
know whether one block is older than another one. This is achieved by
giving each block a time stamp when adding it to the heap. Blocks with
smaller time stamps are older than blocks with bigger time stamps.  This
invariant is preserved across garbage collections (see Section
\ref{subsec:blocks}). Conditional trailing is then implemented
by the following pseudo-code.  \vspace{-1mm}
\begin{verbatim}
   if ((same_block(CellPtr,BH) && (CellPtr < BH))
       || (!same_block(CellPtr,BH) && older_block(CellPtr,BH)))
      trail(CellPtr);
\end{verbatim}
\vspace{-1mm}
If the variable and the heap pointer in the choice point belong to the
same block, then the address order is used; if not, the age of the
blocks is used for comparison.

\subsubsection{Variable binding}
\label{subsubsec:binding}

A common optimisation to reduce trail usage is the following. When a
variable is bound to another variable, the youngest cell is always
bound to the oldest cell. This reduces the chance that the binding
needs to be trailed. This is achieved by variable age testing code
similar to the code for conditional trailing as in Section
\ref{subsubsec:trailing}.

\subsubsection{Tidy trail}
\label{subsubsec:tidytrail}

\Tidytrail is a technique to recover trail space upon cut.  When the
program executes a cut, one or more choice points are discarded. This
can render some entries on the trail useless.  \Tidytrail compacts the
trail by removing these entries. This can reduce the trail size
drastically, for example in benchmark \boyer. \Tidytrail is
particularly important for \incgc, because the trail is part of the
root set, which is scanned at each collection cycle (as will be
discussed in Section \ref{sec:incremental}).  By making the root set
smaller, \tidytrail makes collections cheaper. \Tidytrail was
implemented for both the \optgc and \incgc systems.

\Tidytrail removes trail entries by comparing the ages of the trail
entry and the most recent choice point. The modifications needed are
similar to those made for conditional trailing in Section
\ref{subsubsec:trailing}.

\subsection{Overhead due to heap layout}
\label{subsubsec:overheadhl}

All operations that involve the comparison of cell order (or cell age)
are more expensive in the \newheap. The management of the layout
itself imposes an extra performance cost for \newheap, even if the
program does not trigger garbage collection. In order to measure this
cost, we compare the \oldheap and the \newheap on a number of
benchmarks. Both heap layouts are implemented for hProlog 1.7. The
\oldheap system is always started with a heap big enough to run a
particular benchmark without needing garbage collection.  We
configured the \newheap system for different heap block sizes (the
block size is mentioned in Table \ref{overhead1} in number of heap
cells).  The \newheap systems are started with a heap consisting of
one block. When the \cublock overflows, an extra block is added to the
heap and used for new allocations.  No garbage collection is
performed.  We did not include the benchmarks \mqueens and \chess
here, since they can not run without garbage collection.

\begin{figure}[h]
\begin{centering}
\epsfig{file=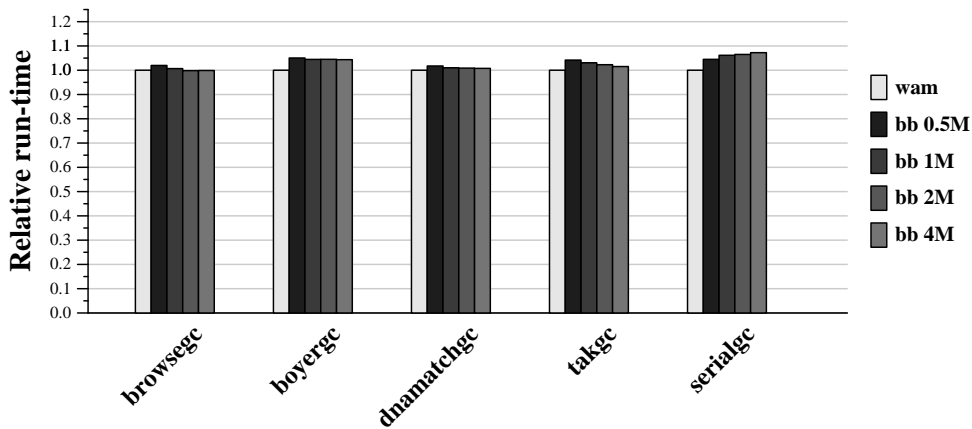,width=5in}
\caption{Overhead of the heap layout}
\label{overhead1_gr}
\end{centering}
\end{figure}

Table \ref{overhead1} shows the time (in ms) needed to run a
particular benchmark with each system.  To isolate the performance
overhead during the execution of a program, we measure the time needed
for running the benchmark only.  I.e., startup time and allocation of
heap space (time spent in malloc and in the UNIX system call mmap) are
not included. The management of new data structures related to the
heap blocks is included in the timings. Figure \ref{overhead1_gr}
shows the performance of the \newheap relative to the \oldheap in a
more graphical way.

\begin{table}[h]
\center
{\small
\input{tn_table1}

}
\caption{Overhead of the heap layout}\label{overhead1}
\end{table}

We observe the following:
\begin{itemize}
\item The benchmarks show that there is some overhead in the \newheap
  systems, but the overhead is in general fairly small with a maximum
  of 7\% for \serial in the \newheap 4M system.
\item For most benchmarks, smaller block sizes in the \newheap systems
  result in slightly more overhead. For a given benchmark, \newheap
  systems with a smaller block size have more blocks; consequently
  there is more overhead due to the switching of \cublock upon
  backtracking. Additionally, there are more \cublock overflows, upon
  which a new block has to be added to the heap.  Finally, the age
  comparison of cells is more expensive globally: the age comparison
  is more expensive when the cells belong to different blocks than
  when they belong to the same block and when the block size is
  smaller, there is a higher chance that the cells belong to different
  blocks.
  
\item For \dnamatch and \boyer, the block size has a much smaller
  influence on the (relative) amount of overhead than for \tak. The
  relative overhead in the \newheap systems depends on the relative
  amount of instructions that become more expensive. The age
  comparison of cells is important here. The age comparison becomes
  more expensive when the cells belong to different blocks, because
  then the time stamps of the corresponding blocks need to be fetched
  and compared.  We measured the total number of age comparisons and
  the relative amount of age comparisons between cells belonging to
  different blocks.  For \tak the relative amount of age comparisons
  between cells belonging to different blocks compared to the total
  amount of comparisons is pretty high (46.5\% in the 0.5M system) and
  is clearly influenced by the block size (from 46.5\% in the 0.5M
  system to 22\% in the 4M system). For \dnamatch and \boyer, on the
  contrary, this number is lower (29.5\% and 29.8\% in the 0.5M
  system) and varies a lot less across different block sizes (from
  29.5\% and 29.8\% in the 0.5M system to 24.2\% and 25.6\% in the 4M
  system). These numbers are in accordance with our explanation.
  
\item \serial runs slower when the block size in the \newheap systems
  becomes bigger.  This goes against our expectation that using
  smaller blocks will result in more overhead.  Since cache effects
  are sometimes the reason for this kind of unexpected results, we ran
  the benchmark with Cachegrind~\cite{NethercotePhD2004}, the cache
  profiler from
  Valgrind~\cite{Sew:valgrind,NethercoteSewardValgrind2003}.  The
  cache measurements indeed confirm that cache effects are at play.
  When the block size is bigger, there are considerably more L2 cache
  misses: 9.5\% more misses for the 4M system relative to the 0.5M
  system.  Most cache misses (around 75\%) happen during unification
  and the difference in cache misses between different block sizes
  comes almost exclusively from a difference in cache misses during
  unification.  The different cache behaviour is caused by the
  different layout of the heap data: because of the small overflow
  space at the end of each heap block, the relative placement of data
  is different across different block sizes.
\end{itemize}

\section{Incremental collection with heap blocks}
\label{sec:incremental}

\subsection{Basic principle} \label{subsec:blocks}

Incremental collection in the \newheap is done by collecting one heap
block during each collection cycle. Since the garbage collector is
based on a copying collector, a \tospace is needed as large as the
\fromspace. This means that one inactive heap block should always be
reserved as \tospace.

\begin{figure}[h]
  \begin{centering}
    \subfigure[Before gc]{\epsfig{file=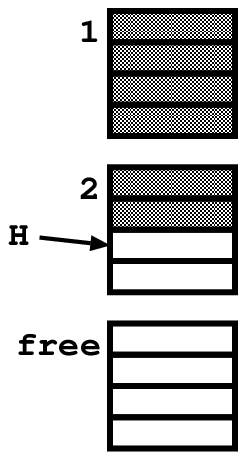,height=2in}%
      \label{blocksA}}
    \subfigure[During gc]{\epsfig{file=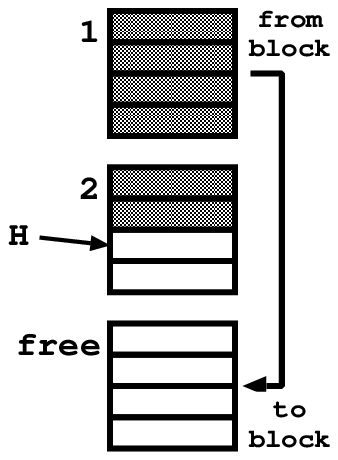,height=2in}%
      \label{blocksB}}
    \subfigure[After gc]{\epsfig{file=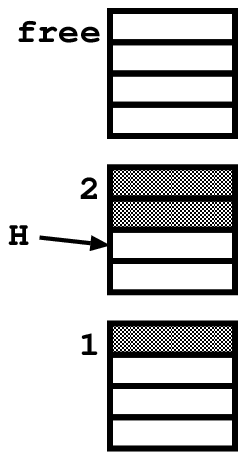,height=2in}%
      \label{blocksC}}
    \caption{Incremental garbage collection in the \newheap}
    \label{blocks}
  \end{centering}
\end{figure}

Figure \ref{blocks} shows what happens during garbage collection.
Before the collection, in Figure \ref{blocksA}, the \newheap consists
of 3 blocks: block 1 (the oldest block), block 2 (the \cublock) and
one block currently not in use. A garbage collection collects one
block (we call this the \fromblock) and copies all its live cells to a
free block (we call this the \toblock).  In this example block 1 is
the \fromblock. During the collection (Figure \ref{blocksB}) block 1
is collected and all its live data is copied to the free block.
Finally in Figure \ref{blocksC} the \toblock becomes the new block
1: it takes the place of the \fromblock in the chain of blocks and
inherits the time stamp of the \fromblock.

Note that if the \fromblock contains garbage (block 1), then the
\toblock still contains some free space at the end of the collection.
Each heap block has a pointer to indicate this.  If the next block
(block 2) is collected in a subsequent collection, its live data can
be copied to this free space (top of block 1). When all free space is
used, the copying is continued in a free block. At the end of
the collection, all (new) blocks used as \toblock are inserted, in
order, into the chain of blocks, right before the \fromblock.  All
inserted blocks receive a time stamp that satisfies the invariant that
older blocks (closer to the beginning of the chain) have smaller
time stamps than younger blocks (closer to the end of the chain). The
\fromblock is removed from the chain.

The free space left in a \toblock can also be used for subsequent
allocation of new data. However, the data allocated in this free space
will inherit the age of the \toblock. In case the \toblock is
different (older) from the \cublock, this can lead to unnecessary
trailing of the newly allocated variables, and it makes instant
reclaiming more difficult. Therefore our allocation policy only
allocates new data in the \toblock if it is the \cublock.

\subsection{Write barriers and remembered sets}
\label{subsec:remsets}

During an incremental collection we want to collect one heap block
independently from the other blocks. During the collection cycle, all
live data found in that block (the \fromblock) is copied to a new
block (the \toblock). The \textit{root set} is defined as a set of
references, which are known to point to live cells in the heap. In the
WAM, the root set consists of references found in the environments,
the choice points, the trail and the argument registers; all these
memory areas contain references to (live) cells on the heap.

\begin{figure}[h]
  \begin{centering}
    \subfigure[Indirect reference]{\epsfig{file=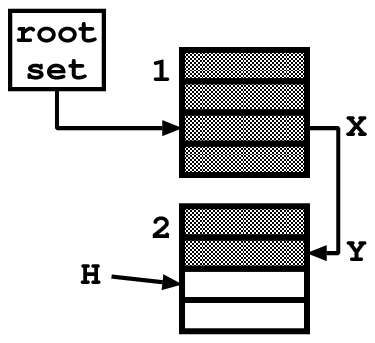,height=1.15in}%
      \hspace{1mm}%
      \label{remsetsA}}
    \subfigure[Write barrier]{\epsfig{file=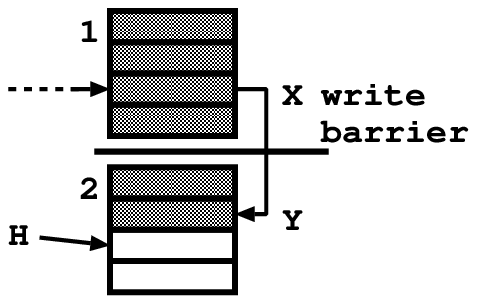,height=1in}%
      \label{remsetsB}}
    \subfigure[Remembered sets]{\epsfig{file=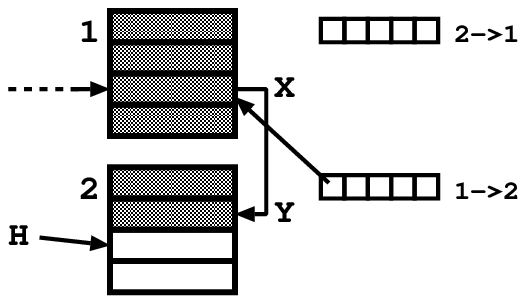,height=1in}%
      \label{remsetsC}}
    \caption{Write barriers and remembered sets}
    \label{remsets}
  \end{centering}
\end{figure}

However, the root set alone is not sufficient to find all live cells
in a heap block without scanning other blocks. In the \newheap, it is
possible that a cell in a block is not referenced by any element of
the root set, although it can be referenced by a live cell in another
block. An example of this is shown in Figure \ref{remsetsA}. Cell
\texttt{Y} is live, because it is indirectly referenced by the root
set, through cell \texttt{X}. At collection time, we want to know
which cells in other blocks have references to cells in the
\fromblock, without scanning all other blocks. References from cells
in one block (the \sourceblock) to cells in another block (the
\targetblock) are called \interblock references; references to cells
in the same block are \intrablock references. A mechanism to remember
the \interblock references is needed. It is common to use a write
barrier and remembered sets as in \cite{wilson92uniprocessor}.

\paragraph{Write barrier}

A write barrier is a mechanism to monitor write operations. Each time
a value is about to be written in a memory cell, this action is
intercepted by the write barrier. The same mechanism is used for
conditional trailing.  The write barrier for trailing checks whether
something is written in a cell older than the current choice point, and
in that case, it puts the relevant cell on the trail. In this
incremental collector, an extra write barrier checks whether an
\interblock reference is created (Figure \ref{remsetsB}) and in that
case, remembers it by recording it in a remembered set (Figure
\ref{remsetsC}).

A more elaborate discussion about write barriers and their
implementation can be found in \cite{JonesLins}. Write barrier
implementations can be software-only or use the virtual memory
hardware (protecting pages for writing). \newheap employs a
software-only write barrier. The code for the write barrier is only
added for assignments where the creation of \textit{inter-block}
references is possible, and has been optimised as described in
\cite{512452}.

\paragraph{Remembered sets}

Remembered sets are collections of \interblock references. These sets
can be organised in several ways. One commonly used option is to have
one remembered set for each block; the write barrier puts all
references to cells in a certain block in that block's remembered
set. Another option is to have remembered sets for each combination of
\sourceblock and \targetblock \cite{beltway02}. Our implementation
uses the latter (Figure \ref{remsetsC}). This configuration has two
important advantages. First, during \gc only the remembered sets that
have the \fromblock as \targetblock need to be scanned. Second, after
\gc, all old entries in the remembered sets should be removed; this
corresponds to removing all remembered sets that have the \fromblock
as either \sourceblock or \targetblock. Note that the write barrier is
also active during \gc: new \interblock references can be created
during \gc and those are added to new or existing remembered sets.

We refer to the \newheap system with the write barrier and \remsets as
the \newheapwb system.

\subsection{Remembered sets and backtracking}
\label{subsec:remsetsbacktrack}

As discussed before in Section \ref{subsec:backtrack}, it is possible
to recover heap space upon backtracking, this is called instant
reclaiming.  All heap cells allocated since the creation of the most
recent choice point can be deallocated and reused for new allocations
after backtracking.  Instant reclaiming however requires special
attention in the presence of \remsets.

\begin{figure}[h]
  \begin{centering}
    \subfigure[Before backtracking]{\epsfig{file=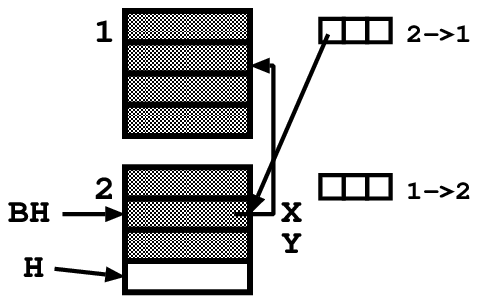,height=1in}%
      \label{remsetbackA}}
    \subfigure[After instant reclaiming]{\epsfig{file=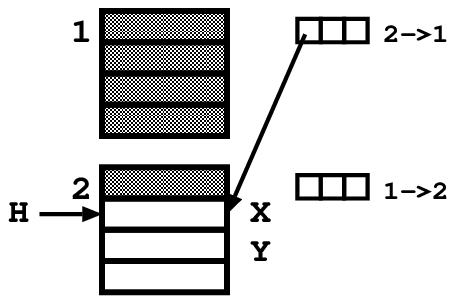,height=1in}%
      \hspace{4mm}%
      \label{remsetbackB}}
    \subfigure[After new allocations]{\epsfig{file=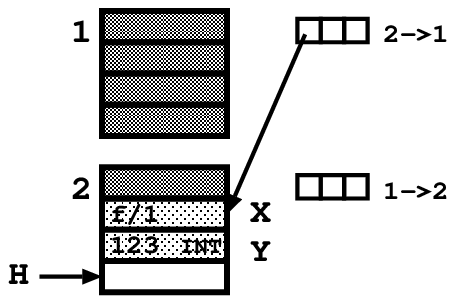,height=1in}%
      \label{remsetbackC}}
    \caption{Remembered sets and backtracking}
    \label{remsetback}
  \end{centering}
\end{figure}

\paragraph{Problem}

It is possible that instant reclaiming deallocates a heap cell, which
is still referenced by an entry in the \remsets. An example of this is
shown in Figure \ref{remsetback}. In Figure \ref{remsetbackA}, before
backtracking, the most recent heap segment contains 2 heap cells:
\texttt{X} and \texttt{Y}. Cell \texttt{X} in \texttt{block 2}
contains an \interblock reference to a cell in \texttt{block 1}, and
thus the \remset (\texttt{2->1}) contains a reference to cell
\texttt{X}.  After backtracking (and instant reclaiming), in Figure
\ref{remsetbackB}, both cells have been deallocated and can be reused
for new allocations. The entry in the \remset (\texttt{2->1}) pointing
to cell \texttt{X} has not been removed. In this case, further
execution allocates the structure \texttt{f(123)}, as seen in
Figure \ref{remsetbackC}.  The \remset entry now points to the cell
containing the untagged header \texttt{f/1}. During a \gc of
\texttt{block 1}, all entries in the \remset (\texttt{2->1}) would be
part of the root set and the untagged value in cell \texttt{X} could
erroneously be interpreted as a valid pointer to \texttt{block 1}.
The cell (apparently) referenced by cell \texttt{X} would be forwarded
and the contents of cell \texttt{X} would be changed accordingly.
Obviously, this is not correct and has to be avoided. A solution for
this problem is to remove all \textit{dangerous} entries (those
referencing untagged heap cells) from the \remsets upon backtracking.

\paragraph{Our solution}

We chose to remove \textit{all} entries in the \remsets that are
younger than the \textit{current} choice point at the time of
backtracking, since those entries are no longer needed.  This ensures
that all \remset entries point to cells inside the \textit{live} heap
(that part of the heap that is currently in use).  This approach has
the additional advantage that upon backtracking some extra memory
space can be freed in the \remsets.

Our technique relies on the fact that during \textit{normal} Prolog
execution (this includes backtracking with instant reclaiming, but
excludes \gc) the order of cells in the heap, the trail and the
\remsets is \textit{chronological}. 

When a variable is bound, the binding is recorded on a \remset if
the binding is across heap block borders (\interblock reference). The
same binding is recorded on the trail if the variable is older
than the \textit{current} choice point. Those conditions are
independent of each other and as a consequence there is no direct
relation between entries on the trail and on the \remsets. The
chronological ordering however guarantees the following invariant:
entries which appear both in the trail and in a particular \remset
have the same relative order in both. This invariant can be used to
recover cells in the \remsets upon backtracking. It leads to the
following rules during untrailing:


\begin{figure}[h]
  \begin{centering}
      \epsfig{file=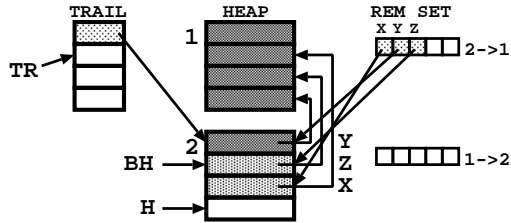,height=1.2in}
    \caption{Removal of remembered set entries upon backtracking}
    \label{remsetbackex}
  \end{centering}
\end{figure}


\begin{itemize}
\item If a trailed cell contains an \interblock reference before
  untrailing, then a reference to this cell should be removed from the
  appropriate \remset. Because of the chronological order in both the
  trail and the \remsets, that particular entry is the most
  recent \textit{trailed} entry in the \remset.
  \begin{list}{}{}
  \item In Figure \ref{remsetbackex}, cell \texttt{Y} is such a
    trailed cell. Upon untrailing, the reference to cell \texttt{Y} in
    the \remset \texttt{2->1} should be removed.
  \end{list}
\item If a trailed reference is removed from a certain \remset, then
  all references added to the \remset after the trailed reference,
  should also be removed.
  \begin{list}{}{}
  \item The trailed cell \texttt{Y} has been removed from \remset
    \texttt{2->1}. Also the more recent entry, the reference to cell
    \texttt{Z}, should be removed.
  \end{list}
\item After untrailing and applying the above rules, a number of
  \remsets might still contain some entries which are younger than the
  \textit{current} choice point. Since the bindings recorded by those
  entries were not trailed (otherwise they would have been removed
  during untrailing), those entries refer to heap cells which are
  younger than the \textit{current} choice point and are part of the
  most recent heap segment which will be removed through
  \textit{instant reclaiming}.  So, after untrailing, all top most
  entries in the \remsets which point to cells younger than the
  \textit{current} choice point should be removed.
  \begin{list}{}{}
  \item Scanning the \remsets reveals that the \remset
    \texttt{2->1} contains a reference to cell \texttt{X}, which
    belongs to the most recent segment and thus should be removed.
  \end{list}
\end{itemize}

\paragraph{Garbage Collection}

During a \gc, the live cells in the \fromblock are copied to the free
space in one (or more) empty or partially filled heap block(s) (the
\toblock). During the copying, newly created \interblock references
are added to the appropriate \remsets.  This has been discussed in
Section \ref{subsec:remsets}.

The order in which the live cells are copied, determines which \remset
entries are created and in what order. After collection, the order of
the new entries in the \remsets will be the same as the order of the
corresponding cells in the \toblock. Since the copying order is not
known, no assumptions can be made about the order of the new \remset
entries. More importantly, it is no longer guaranteed that the order
in the \remsets is \textit{chronological}, i.e. corresponding to the
trail order.

Because of this, the aforementioned set of rules to remove \remset
entries upon backtracking is not valid after \gc. It is important to
note however, that the problem with backtracking, as described before,
does not exist in that part of the heap that has been garbage
collected (and the corresponding part in the \remsets).  Namely, the
problem only appears when \textit{instant reclaiming} causes some
entries in the \remsets to point to freed heap space, but since the
used copying algorithm does not preserve heap segments,
\textit{instant reclaiming} is not possible on collected parts of the
heap.

Since there is no problem upon backtracking to a \textit{collected}
choice point, there is no need to remove the related \remset entries.
All \remset entries created during a \gc are tagged as \texttt{GC}, so
they can be identified as such.  Still, when subsequent backtracking
reclaims a collected part of the heap, the relevant \texttt{GC}-tagged
entries should be removed. We call the top of the copied heap cells
after a \gc the \textit{garbage collection top}. A collected part of
the heap is reclaimed when heap space is reclaimed beyond its
corresponding \textit{garbage collection top}.  After a collection, an
extra cell is added to all \remsets with new entries.  This extra cell
holds a reference to the \textit{garbage collection top} and is tagged
as \texttt{GCT}. The \texttt{GCT}-cell indicates when the accompanying
\texttt{GC}-cells in the \remset should be removed.

This results in two extra rules for removing \remset entries upon
backtracking:

\begin{itemize}
\item If the most recent entry in a \remset is \texttt{GCT}-tagged
  (\textit{garbage collection top}), then don't remove any cells from
  that \remset.
\item After untrailing, all relevant \remsets should be scanned: if
  the most recent entry is \texttt{GCT}-tagged, then check whether
  backtracking was beyond the referenced heap cell and if so, remove
  all entries on the top of the \remset which are \texttt{GC}-tagged.
\end{itemize}

\paragraph{Optimisations}

There are two important optimisations to this approach. Upon
backtracking, only the \textit{relevant} \remsets should be scanned
and in a very common situation even none at all.

There is no need to check \textit{all} \remsets upon backtracking.
Because of instant reclaiming, the top of the heap is reset upon
backtracking. We refer to the heap block containing this \textit{new}
heap top as the \textit{backtrack block}. Only those \remsets should
be scanned that have the \textit{backtrack block} as \sourceblock.
These are the only ones that can contain entries which have not been
trailed.

Investigating the execution of several Prolog programs revealed that
in most cases of backtracking, no new heap cells are created between
choice point creation and backtracking. This means that all \interblock
references that might be created, are trailed. Consequently, there is
no need to scan the \remsets after untrailing, when backtracking did
not recover any heap space.

\paragraph{Additional remarks}

The solution we presented, seems rather complicated and one might
think it would be easier to use the trail to store all \interblock
references.  Upon backtracking and instant reclaiming, it would be
easy to remove the appropriate \interblock references.  Still, using
the trail for storing both references which have to be reset upon
backtracking and \interblock references has important disadvantages.
An immediate drawback is that the whole trail and thus \textit{all}
\interblock references will need to be scanned during a collection
cycle.  However, there is an even worse issue, which is caused by the
unknown relation between \interblock references before and after a
collection cycle.  More specifically, during the collection, it is
possible that new \interblock references are created and that some
disappear.  The new \interblock references would need to be inserted
in the appropriate trail segment (and not at the top of the trail), so
as to make sure that they are released when needed and not too early
or too late.  The issues related to keeping the trail consistent make
that approach unattractive and as complex as our current approach.

Another approach we have considered, is to trail the remembered set
entries themselves.  Upon backtracking, the remembered set entries
could then be reset to \textit{nil}.  Also this approach has important
drawbacks.  For each \interblock reference two extra references would
be created: one in the remembered sets and one on the trail.
Additionally, the remembered sets would need to be scanned and cleaned
up regularly.  Finally, this solution would have similar issues as the
previously mentioned one: references to \interblock references created
during a collection, must be added to the appropriate trail segment.

\subsection{Partial early reset}
\label{subsec:partialearlyreset}

During garbage collection, it is possible to reset certain bindings
recorded on the trail (or for a value-trail to reset the recorded
cells to the value they would get upon backtracking); this is called
\earlyreset. With \earlyreset, the garbage collector takes over some
of the untrailing work that would be done upon backtracking. This is
beneficial for the same reason as \tidytrail. Trail entries, subject
to \earlyreset, can be removed from the trail, which can lead to a
smaller trail and thus a smaller root set.

\Earlyreset is only possible for cells referenced by the trail, which
are not used in the forward execution of the program. For each
choice point, for which we can compute the forward execution and the
corresponding set of reachable data, we can apply \earlyreset on the
accompanying trail segment (that part of the trail that would be
untrailed upon backtracking to that specific choice point). It is
possible to determine the reachable data from a certain choice point
and apply \earlyreset to the appropriate trail entries during garbage
collection. A more in-depth discussion about this technique can be
found in \cite{SicstusGarbage@CACM-88,BekkersRidouxUngaro@IWMM-92}.

\Earlyreset relies on the fact that, during a \gc, we can find all
live data in the forward execution of each choice point. This is done
by recursively marking or forwarding all cells from the root set which
are reachable through the environment of a certain choice point. Cells,
that are not reachable through that environment are eligible for
\earlyreset. Those cells will not be used in the forward execution and
can safely be reset to the value they will get upon backtracking.

\begin{figure}[h]
\begin{centering}
\epsfig{file=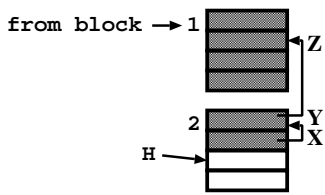,height=1in}
\caption{Indirect reference}
\label{tn_partialer_gr}
\end{centering}
\end{figure}

In a \newheapwb system, with incremental collections, \earlyreset is
not possible without some modifications. To find all cells reachable
in the forward execution, also cells in other heap blocks than the
\fromblock (the heap block that is being collected) should be scanned.
This is needed because reachable cells in other heap blocks might have
(indirect) references to cells in the \fromblock, as shown in Figure
\ref{tn_partialer_gr}. In this case, cell A is reachable in the
forward execution. Cell A keeps cell B live and cell B keeps cell C
live. Although there is no direct reference to cell C in the forward
execution it can be live because of references from other heap blocks.
Although \earlyreset is not possible without looking at other heap
blocks, an approximation of \earlyreset is possible. Normal
\earlyreset is not possible because of reachable cells in other heap
blocks that have references to the \fromblock. All those cells,
however, can be found in the \remsets; as discussed in
\ref{subsec:remsets}, the \remsets contain all \interblock references.

The approximation, which we call \partialer, is a conservative
approximation of \earlyreset and assumes that all references from
other heap blocks to the \fromblock are live in the forward execution.
Before scanning cells in the reachable environments, forwarding is
first started from references found in the \remsets. Note that (at
this moment) only those remembered sets need to be scanned, that have
the \fromblock as \targetblock. After forwarding those cells
referenced through the remembered sets, \earlyreset can be done as
with a standard heap layout. However, because \partialer is less
precise (more conservative), and because it can only be done for cells
in the \fromblock, it recovers less space on the trail than standard
\earlyreset.

For the implementation of \partialer, one more change is needed in the
basic \gc algorithm. \Earlyreset makes it possible to compact the
trail during the \gc; but for the actual compaction of the trail the
choice point stack needs to be traversed twice instead of only once.
The first pass over the choice point stack is a pass from new to old
and is used to forward cells from the reachable environment cell in
the forward continuation from a certain choice point. During this first
pass, all trail cells that can be removed due to \partialer are
marked. Also, the links between consecutive choice points are reversed.
In the second pass, the choice point stack (and the trail stack) are
traversed from old to new. During this pass, the choice point links
are restored again, the trail is compacted and the trail pointers in
the choice points adapted accordingly.

We measured the effectiveness and the overhead of \partialer in the
final system, the incremental collector, which will be discussed in
Section~\ref{sec:policies}.  There is an extra cost in collection time
because of \partialer, since the choice point stack needs to be
traversed twice.  Still, the overhead is reasonable and always lower
than 10\%.  There is one exception though: for the benchmark \chess,
the collection time increases 50\% when \partialer is applied.  The
reason is twofold.  First, as can be seen in Table~\ref{benchmarks},
\chess has a large root set.  \chess has a large choice point stack,
which needs to be traversed twice during each collection and this
leads to a higher extra cost.  Second, when \partialer is used, the
benchmark needs less memory (4\%) to run, but this results in more
collections (51 instead of 36).  We believe this is an exceptional
case: when \partialer is used, the savings in heap space are just big
enough to not always expand the heap, when it would be expanded
if \partialer were not used.  The effectiveness of \partialer is
visible in the trail size needed to run the benchmark (the high
watermark of the trail).  There are also savings in heap space, which
sometimes result in fewer collections, but the savings are in general
very small (less than 1\%) and almost negligible.  The reduction in
needed trail size is dependent on the benchmark (the reduction can be
as high as 80\%), but is in line with the effects of normal early
reset in the standard optimistic collector.


\subsection{Overhead due to write barriers and remembered sets}
\label{subsec:wbremset}

To investigate the effect of the write barrier on program run-time, we
compare the \oldheap, the \newheap (without write barrier) and the
\newheapwb (\newheap with write barrier and remembered sets). Again we
present \newheapwb systems for different sizes of the heap blocks. The
same settings apply as in Section \ref{subsubsec:overheadhl}. None of
the systems perform garbage collection during the benchmark runs.

\begin{figure}[h]
\begin{centering}
\epsfig{file=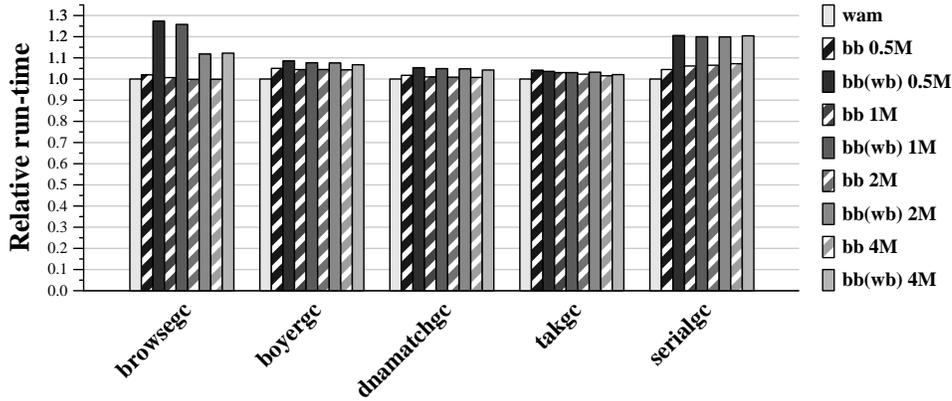,width=5in}
\caption{Overhead of the write barrier and remembered sets}
\label{overhead2_gr}
\end{centering}
\end{figure}

Table \ref{overhead2} contains the timings for the benchmarks for each
\newheapwb system \textbf{t}$_{\bf{tot}}$ (in ms). The maximal size of
the remembered sets \textbf{m}$_{\bf{rs}}$ (number of entries) is also
included. Figure \ref{overhead2_gr} shows the performance of the
\newheapwb systems relative to the performance of the \oldheap system.
The graph also includes the timings for the \newheap systems, as
presented in \ref{subsubsec:overheadhl}.


\begin{table}[h]
\center
{\small
\input{tn_table2}

}
\caption{Overhead of the write barrier and remembered sets}\label{overhead2}
\end{table}


From the figures we observe that:
\begin{itemize}
\item The write barrier gives an extra overhead; the total overhead of
  the \newheapwb systems compared to the \oldheap system can be as
  high as 27\% (\browse in \newheapwb 0.5M).
\item Benchmarks (\boyer, \dnamatch, \tak), where most references are
  very local (the reference and the data it points to are very close
  to each other in the heap), experience a rather small overhead from
  the write barrier: in these cases the total overhead of \newheapwb
  remains lower than 10\% compared to \newheap. The reason is that
  although the write barrier is triggered, it rarely needs to add
  references to the remembered sets. The measurement of the maximal
  size of the \remsets as seen in Table \ref{overhead2} is in
  accordance with this explanation.
\item \serial has a much higher overhead in the \newheapwb systems.
  The amount of overhead is more or less the same for all tested
  \newheapwb systems and is not affected by the block size. In the
  first phase of this benchmark, some data structures are initialised
  and then, in the second phase, a tree is built with references to
  these structures. So, all newer data contains references to the
  oldest data on the heap. This results in a lot of write barrier
  activity and compared to \boyer, \dnamatch and \tak the write
  barrier traps a higher percentage of \interblock references, which
  have to be inserted into the remembered sets. The high amount of
  write barrier activity can also be deduced from the very high number
  of \remset entries.
\item \browse is a benchmark where most references are very local,
  still it experiences very high overhead from the write barrier.
  Profiling revealed that in this benchmark, upon backtracking and
  untrailing it is often needed to remove several cells at once from a
  certain remembered set. Since this is costlier, it might explain the
  higher overhead (compared to \serial where this does not happen).
  The overhead is lower and almost the same for the \newheapwb 2M and
  4M systems.  The reason is that the benchmark needs only 1.5M heap
  space; it uses only one heap block in the 2M and 4M systems.
\item The size of the remembered sets becomes smaller as the block
  size increases. This is what we intuitively expect, because moving
  from small blocks to bigger blocks, some \interblock references
  become \intrablock references and as such, there will be less
  entries in the remembered sets. This can clearly be seen in \browse,
  \boyer and \dnamatch. For \serial, the number of \interblock
  references is nearly the same for all blocks sizes; as mentioned
  earlier, the reason is that most references in the heap point to
  data at the bottom of the heap.
\end{itemize}

\section{Garbage collection policies}
\label{sec:policies}

The \gc policy defines which events trigger a collection, and
which part of the heap is collected when a collection is
triggered. The basic \newheap framework allows for many different \gc
policies. Several events can be used as triggers for \gc,
such as overflow of the \cublock, a \remset exceeding a certain size,
a certain number of allocated blocks, \ldots.

In the two configurations that we present, the only trigger we use, is
the number of newly allocated heap cells since the previous \gc. This
trigger allows us to impose a certain frequency of collections as a
function of the amount of allocated heap space. We use it to schedule
two garbage collections for each new heap block allocated, to ensure
that \gc keeps up with new allocations.

In the following sections, we discuss two policies: a purely
incremental policy and a generational policy.

\subsection{Incremental policy}
\label{subsec:incpol}

A garbage collector with an incremental \gc policy collects the heap
incrementally with the purpose of having smaller pause times.
Commonly, this is done by dividing the heap in small increments, which
are collected one at a time, independently from each other.  This has
the purpose of amortising the total cost of collecting the heap:
instead of one large pause time (collecting the whole heap at once),
there are multiple smaller pause times (gradually collecting the whole
heap by collecting one increment at a time).  For several
applications, such as interactive applications and control systems,
small pause times are desirable and sometimes even demanded.

In incremental collectors, it is important that the collection of used
heap space makes enough progress compared to the allocation of new
heap space. If allocation is faster than collection, the program will
eventually run out of memory.

In the \newheapwb systems, the natural choice for an increment, is a
heap block. We decide to collect one increment (heap block) during
each collection. To make collection keep up with allocation, we chose
to schedule a \gc after half a heap block has been allocated.  The
heap blocks are collected in a round-robin fashion.  This guarantees
that over time, each heap block will be subject to collection, thus
ensuring that the whole heap is gradually collected.

As a fine-tuning, the collector also takes into account the
\textit{survival rate} of a heap block. The survival rate of a heap
block is used as a heuristic for the ratio of cells that will survive
the next \gc of that block. Initially, when a free block has been used
as \toblock for the first time, the survival rate is computed as
follows:

\[rate_{surv} = \frac{block_{live}}{block_{total}}\]

In this formula, \textit{block$_{live}$} is the number of cells that
survived this collection and \textit{block$_{total}$} is the total
number of heap cells in the \fromblock. Each heap block has a set of
\textit{block$_{live}$} and \textit{block$_{total}$} values. After a
\gc, those values are updated by adding the new
\textit{block$_{live}$} and \textit{block$_{total}$} values to the old
values stored in the heap block.

It is possible that during a collection, the live cells from the
\fromblock are copied to more than one heap block. In that case, the
value \textit{frac$_{live}$} for such a block is the relative amount
of cells that have been copied to that block:

\[frac_{live} = \frac{live_{block}}{live_{total}} \]

The value \textit{live$_{block}$} is the number of live cells that
have been copied to this heap block and the value
\textit{live$_{total}$} is the total number of heap cells that
survived the \gc.

For a heap block with the values \textit{block$_{live}$} and
\textit{block$_{total}$} before collection, the survival rate after
collection is defined as follows:

\[rate_{surv}' = \frac{block_{live}'}{block_{total}'} =
\frac{block_{live} + frac_{live} \cdot live_{total}}{block_{total} +
  frac_{live} \cdot gc_{total}}\]

\textit{gc$_{total}$} is the total number of heap cells in the
\fromblock.

The survival rate of a heap block is used as a heuristic for the
amount of heap cells that will survive during the next collection of
that heap block. When the survival rate of a heap block is high, it is
probably not worthwhile to collect that particular block. Since the
survival rate is high, it is assumed that a collection would only
recover a small amount of heap space and thus it seems better to
collect another heap block. The garbage collector skips heap blocks
with a survival rate of more than 80\%. However, under these
conditions, a skipped heap block will never be collected again. To
guarantee new collections at some point, the survival rate of a
skipped heap block is diminished by 10\% (by adapting the
\textit{block$_{live}$} value).

\begin{figure}[h]
\begin{centering}
\epsfig{file=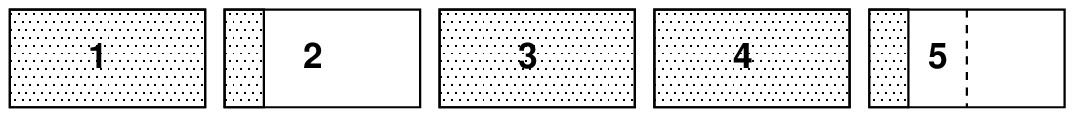,width=4in}
\caption{Incremental heap configuration}
\label{inc_cfg}
\end{centering}
\end{figure}

Figure \ref{inc_cfg} shows an example heap configuration for the
incremental collector. The heap consists of five heap blocks, with 1
the oldest and 5 the most recent block. The most recent block is
divided in two halves and each time one half is filled, a \gc is
triggered. During the collections heap blocks 1-4 are collected in a
round-robin fashion; this happens in the order from old to new. The
most recent heap block, block 5, is only collected when it is full. In
this particular configuration, block 1 and 2 have already been
collected.  During the next collection, block 3 is collected and
the live cells from block 3 are copied to the remaining free space
in block 2. When more space is needed during the collection, a new
empty heap block is inserted at that place in the heap.

\subsection{Generational policy}
\label{subsec:genpol}

A generational garbage collector takes into account the age of data:
it frequently collects the young data and much less often the old
data. This policy is based on the \textit{generational hypothesis},
which assumes that in most cases young data dies soon, while older
data lives much longer \cite{JonesLins}.  Generational collectors
typically divide their heap into two or more spaces (also called
\textit{generations}) to segregate data by age.  New allocations are
always made in a special space, called the nursery space.  Whenever
data in the nursery space survives a garbage collection, it is
promoted to another space containing older data.  Apart from the
nursery space, a generational collector can have any number of older
spaces which contain data with varying degrees of age.  The older the
space, the less frequently it is collected.  Whenever a space is
collected, also the younger spaces are collected at the same time.

Traditionally, generational \gc aims to reduce the total time spent on
\gc by focusing collection efforts on those parts of the heap where
collection will presumably be the most effective (recover the most
space).  We use the same idea that generational \gc is based on (the
generational hypothesis) to implement a more complex policy for the
incremental collector.%
\footnote{The use of the term \textit{generational} to refer to this
  policy and the collector using it might be misleading, since we do
  not use the term in its traditional meaning.  We use the term to
  make it clear that the policy is based on generational \gc and to
  make the distinction with the simple incremental policy discussed
  before.}
We believe such a policy will be more effective than the simple
incremental policy discussed before (in Section~\ref{subsec:incpol})
for programs where the generational hypothesis holds.

\begin{figure}[h]
\begin{centering}
\epsfig{file=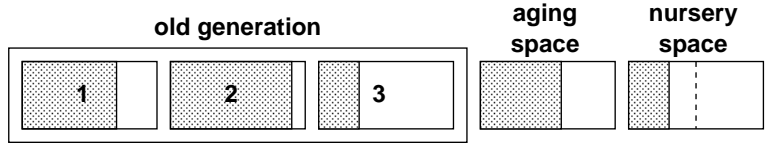,width=4in}
\caption{Generational heap configuration}
\label{gen_cfg}
\end{centering}
\end{figure}

In our implementation, we divide the heap into three spaces (see
Figure \ref{gen_cfg}): a nursery space, an aging space and the old
generation. Both the nursery and the aging space are fixed size and
consist of one heap block each, while the old generation is expanded
with new heap blocks as needed. New allocations are always done in the
nursery space, which contains the top of the heap. A \gc is triggered
when (half of) the nursery space overflows. When that happens, all
live cells from the nursery space are copied to the aging space.  The
aging space is used as a buffer between the nursery space and the old
generation; it also guarantees that all data ending up in the old
generation have survived at least two collections. When the aging
space overflows during a \gc of the nursery space, a collection of the
aging space is scheduled. During a \gc of the aging space, the live
cells in the aging space are copied to the top of the old generation.
Whenever the old generation is expanded with a new heap block, the
following collections alternately collect a heap block from the old
generation and the nursery space till all blocks in the old generation
have been collected. This is done to ensure that heap blocks in the
old generation are subject to collection at some point in time.  Upon
backtracking, it is possible that heap space can be recovered in the
aging space and even in the old generation through instant reclaiming.
Although heap space will be recovered in those areas, new allocations
after instant reclaiming are always done in the nursery space. As a
fine-tuning, collections of heap blocks belonging to the old
generation also take into account the \textit{survival rate} of the
collected block. This is similar to the one for the incremental
collector, discussed in the previous section.

An example configuration can be seen in Figure \ref{gen_cfg}. In this
configuration, the last collection was a collection of the nursery
space. In the meantime, new allocations have been done. When half of
the nursery space is filled, a new collection is triggered. During
this collection the next block (block 2) of the old generation is
collected, and its live cells are copied to the previous block in the
old generation. After the collection, the execution of the program
continues till the nursery space overflows. In the ensuing \gc the
nursery space will be collected.

\subsection{Remarks and ideas for future work}
\label{subsec:rempol}

An interesting question is whether \gc can \emph{reduce} the number of
\remset entries.  During the collection of a heap block, several
effects influence the number of \interblock references.  Most probably
some of the data in the heap block will not survive the collection and
consequently a number of \interblock references (and thus \remset
entries) might disappear. As explained in Section~\ref{subsec:blocks},
the live data of the collected block is copied to the remaining free
space in the previous block.  This means that the \interblock
references between the collected block and the previous block will
become \intrablock references and thus disappear from the \remsets.
Still, it is possible that the live data does not fit in the remaining
free space of the previous block and that part of the live data ends
up in the previous block and the rest in another (new) block.  In this
case, it is possible that some \intrablock references become
\interblock references and thus, that the number of \remset entries
increases.  During \gc, the previously mentioned effects are all at
play and the global effect can either be an increase or a decrease of
the number of \remset entries. The only way to guarantee that this
number does not increase during a collection, is to enforce that the
live data of a block always ends up in one and the same block after a
collection.

In the current implementation, all heap blocks have the same fixed
size.  We have taken this approach, since it allows us to easily (and
at a constant cost) find the heap block a particular cell belongs to.
Nevertheless, it is possible to build a similar system based on the
same ideas, which allows for a heap with blocks of different sizes
(like in the Oz implementation).  This is worth investigating, since
it could be used to dynamically tune the pause time.  For example, a
policy could be implemented that when the survival rate in a heap
block is too high, the surviving data is copied to smaller blocks
during a collection, with the goal of reducing the pause time upon
collecting the same data again.  When different block sizes are
allowed in the same heap, the heap block(s) which typically have a low
survival rate (the heap block on the top of the heap that is used for
new allocations) could be made larger.

The incremental collector, as described here, collects only one heap
block at a time.  It seems worthwhile to investigate collecting two
(or more) heap blocks at once.  When two heap blocks are collected at
the same time, there is no need to scan the \remsets which contain
references between the two blocks.  This would allow to schedule the
concurrent collection of two heap blocks which have many \interblock
references.

\section{Experimental results}
\label{sec:experiments}

We implemented the two policies, as discussed in Section
\ref{sec:policies}, in the \newheapwb system. We call the collector
with the incremental policy \incgc, and the collector with the
generational policy \gengc. Both collectors are implemented for
hProlog and use \textit{optimistic copying} as underlying algorithm.
We compare several \incgc and \gengc collectors (with varying block
sizes) with the standard hProlog 1.7 collector, \optgc.  The \optgc
system is always started with initial heap size of 2M heap cells. The
garbage collection policy in \optgc is as follows: a garbage
collection is triggered whenever the heap is full, all collections are
\textit{major} collections (they collect the whole heap at once) and
\optgc doubles the heap size whenever more than 70\% of the heap
survives a collection.

Tables \ref{incgctable} and \ref{gengctable} contain the results of
benchmarks with \incgc and \gengc. We do not present the benchmarks
\browse, \dnamatch and \tak here; they were left out because they are
too small to be interesting concerning heap usage. We
measured the following items:
\begin{myitemize}
\item \textbf{t}$_{\bf{gc}}$ total time spent on garbage collection
\item \textbf{t}$_{\bf{tot}}$ total run-time, including t$_{gc}$
\item \textbf{n}$_{\bf{gc}}$ number of garbage collections
\item \textbf{t}$_{\bf{l}}$ minimum pause time (time needed for one
  garbage collection cycle)
\item \textbf{t}$_{\bf{a}}$ average pause time
\item \textbf{t}$_{\bf{u}}$ maximum pause time
\item \textbf{m}$_{\bf{alloc}}$ total allocated heap space, including
  the \tospace
\item \textbf{m}$_{\bf{hp}}$ high watermark memory usage in the heap
\item \textbf{m}$_{\bf{rs}}$ high watermark memory usage in the
  remembered sets
\end{myitemize}
Time measurements are in milliseconds, space measurements in heap
cells. For \incgc and \gengc the total allocated heap space
\textbf{m}$_{\bf{alloc}}$ is indicated as $x/y$ with $x$ the number of
blocks and $y$ the number of cells.  Note the difference between
\textbf{m}$_{\bf{alloc}}$ and \textbf{m}$_{\bf{hp}}$:
\textbf{m}$_{\bf{alloc}}$ is the amount of memory on the heap that is
potentially needed during a \gc, while \textbf{m}$_{\bf{hp}}$ is the
highest usage of heap space during normal execution of the benchmark
(i.e., excluding collections).

\subsection{Experimental results of the incremental collector}
\label{subsec:resultinc}

\begin{table}[h]
\center
{\scriptsize
\input{tn_inc_table}

}
\caption{Benchmark results from the incremental collector}\label{incgctable}
\end{table}

From the benchmark results of the incremental collector, we observe
the following:
\begin{itemize}
\item The average pause time in the \incgc systems is always smaller
  than in \optgc. Within the \incgc systems, smaller block sizes
  result in smaller pause times. The reason for this is that the
  amount of work for the incremental collector increases with the
  block size. The relative difference between pause times across
  different block sizes depends on the size of the root set; the
  relative difference becomes smaller as the root set becomes bigger.
  This can be seen when comparing \mqueens and \chess. In the case of
  \mqueens the root set is very small (see Table \ref{benchmarks}) and
  the average pause time almost doubles as the block size doubles; in
  the case of \chess the root set is very large (see Table
  \ref{benchmarks}) and the relative difference of the pause times
  across different block sizes is much smaller.
\item The number of garbage collections in \optgc is in most cases
  (much) lower than in \incgc. This is to be expected since the
  collections in \optgc are \textit{major} and collect the whole heap,
  while the collections in \incgc are \textit{minor} and collect only
  a (small) part of the heap. For the \incgc systems, there are also
  more collections as the block size goes down.
\item Total garbage collection time in \incgc is in most cases bigger,
  but sometimes smaller than in \optgc. A priori, we expect garbage
  collection time in \incgc systems to be higher, since they have more
  collections and scan the root set more often than \optgc (during the
  benchmark). This can clearly be seen in the case of \chess, where
  the difference in total garbage collection time between \incgc 0.5M
  and \optgc is a factor 4.4.  However, it is difficult to draw
  conclusions from these timings. First of all, garbage collection in
  different systems is triggered at different times. And second,
  \incgc and \optgc use a different garbage collection policy; which
  policy is best depends heavily on the characteristics of the
  benchmark. Within the \incgc systems, we see that the total garbage
  collection time is smaller for bigger block sizes. This can be expected
  since smaller block sizes need more collections and
  consequently the root set is scanned more often.
\item In most cases, the total amount of allocated memory for the heap
  (\textbf{m}$_{\bf{alloc}}$) is lower for the \incgc systems than for
  the \optgc system.  This is what we expected because of the smaller
  \tospace. In some cases (\chess in \incgc 2M and 4M) \incgc needs
  more memory. One reason for this is that \incgc keeps more cells
  live than \optgc, because it supposes that all cells in the
  remembered sets are live, while this is not always true.
\item The total amount of used heap space (\textbf{m}$_{\bf{hp}}$)
  is in many cases smaller in the \incgc systems than in the \optgc
  system.  This is mainly a consequence of the difference in expansion
  policies between the two systems: in the \optgc system, the heap
  size is doubled upon expansion, while in the \incgc system an extra
  heap block is added to the heap.  For example, if the benchmark
  needs a minimal heap size of 5M cells to run with \gc, this will
  result in a heap of 8M cells for the \optgc system, while the 2M
  \incgc system could run the benchmark with a heap of 6M cells (3
  blocks of 2M).
\item The relative amount of free space
  (\textbf{m}$_{\bf{alloc}}$-\textbf{m}$_{\bf{hp}}$) in the heap
  blocks is reasonable for most benchmarks: it is around or below 20\%
  for \boyer, \serial and \mqueens.  The ratio is however bigger for
  \chess: from 25\% in the 0.5M system up to 50\% in the 4M system. In
  general, the ratio of free space is bigger when the block size is
  bigger.  The reason is that it is possible that one heap block,
  which has been used as \toblock, is nearly empty (like block 2 in
  Figure~\ref{inc_cfg}).  When the block size is bigger, the total
  number of blocks in the heap is smaller and the impact of one such
  block on the ratio of free space will be bigger.
\item The size of the remembered sets (\textbf{m}$_{\bf{rs}}$) depends
  on the characteristics of the benchmark.  For \mqueens and \boyer
  the memory usage of the remembered sets is small (less than 1\%)
  compared to the memory usage of the heap. This number is noticeably
  higher for \chess (up to 17\%) and \serial (up to 37\%). The number
  of \interblock references is very high in \serial, because most
  references created during the benchmark refer to cells at the bottom
  of the heap.
\item In the \incgc systems, in many cases the remembered sets become
  smaller as the block size increases.  As already noticed in Section
  \ref{subsec:wbremset}, this is because moving from small blocks to
  bigger blocks, some \interblock references can become \intrablock
  references.  However, this reasoning is only valid in the absence of
  garbage collection.  As soon as garbage collection comes into play,
  one has to take into account that systems with different block sizes
  trigger collections at different moments.  Consequently, data may be
  placed in unrelated locations on the heap.  For example, a tree data
  structure may be placed within one small block in one system, while
  it is spread over two large heap blocks in another system, which
  leads to more \interblock references in the system with the larger
  block size.
\item The total run-time in \incgc systems with small block sizes is
  always bigger than in \optgc. This is partly due to the overhead at
  run-time, and partly due to the bigger total garbage collection time
  as already mentioned.
\end{itemize}

The benchmarks show that the incremental garbage collector, \incgc,
results in substantially lower pause times, and that it
generally has a lower memory usage than \optgc.

\subsection{Experimental results of the generational collector}
\label{subsec:resultgen}

\begin{table}[h]
\center
{\scriptsize
\input{tn_gen_table}

}
\caption{Benchmark results from the generational collector}\label{gengctable}
\end{table}

Most of the points from the discussion of the results of the
incremental collector, are also valid for the generational collector.
From the results of the generational collector, we observe the
following:
\begin{itemize}
\item The average pause time in the \gengc systems is always smaller
  than in \optgc. See remarks in Section \ref{subsec:resultinc}.
\item The number of garbage collections in \optgc is in most cases
  (much) lower than in \gengc. See remarks in Section
  \ref{subsec:resultinc}.
\item There are no consistent differences in the number of garbage
  collections between the \gengc and the \incgc systems, except for
  the benchmark \mqueens. The \mqueens benchmark performs very well
  with a generational policy and the number of garbage collections in
  the \gengc systems is a lot less than the number of garbage
  collections in the \incgc systems with the same block size.
\item Total garbage collection time in \gengc is for \chess and
  \serial larger, and for \mqueens and \boyer smaller than in \optgc.
  The reason is that \mqueens and \boyer are better suited for
  generational policies.
\item In most cases, the \gengc systems have a lower memory usage
  (\textbf{m}$_{\bf{alloc}}$) than the \optgc system. See remarks in
  Section \ref{subsec:resultinc}.
\item The total amount of used heap space (\textbf{m}$_{\bf{hp}}$) is
  in many cases smaller in the \gengc systems than in the \optgc
  system. See remarks in Section \ref{subsec:resultinc}.
\item The relative amount of free space in the heap blocks is bigger
  in the \gengc systems than in the \incgc systems.  The reason is the
  division of data in generations. For each generation, it is possible
  that it contains one block which is almost empty; this can be seen
  in Figure~\ref{gen_cfg}. Especially in the systems with bigger block
  sizes --and thus fewer blocks-- such an almost empty block will have
  a bigger impact on the average amount of free space in a heap block.
\item The \gengc systems have a higher memory usage than the \incgc
  systems for most benchmarks. The reason is that in the \gengc
  systems the old generation is collected less often and might contain
  a larger amount of no longer needed data. The heap blocks in the \incgc
  system are collected in a round-robin fashion, and older blocks are
  collected more frequently than the heap blocks in the old generation
  of the \gengc systems.
\item The remembered sets in the \gengc systems are generally smaller
  when block sizes are bigger. This can be expected since moving from
  small blocks to bigger blocks, some \interblock references 
  become \intrablock references. However, the remarks about this as
  given in Section \ref{subsec:resultinc}, are also valid here.  There
  is no obvious relation between the size of the remembered sets in
  \incgc and \gengc systems.
\item The total run-time in \gengc systems is in most cases longer
  than in the \optgc system. The exception is \mqueens, where all
  \gengc systems perform better than the \optgc system. Both \boyer
  and \mqueens perform well with a generational garbage collector, but
  in the case of \mqueens the difference in garbage collection time is
  is big enough to result in better total run-time. For \boyer, the
  improvement in garbage collection time is smaller than the overhead
  due to the \newheapwb system.
\end{itemize}

The benchmarks show that the incremental garbage collector, \gengc,
succeeds to obtain lower pause times, and that in most cases it has a
lower memory usage than \optgc. For some benchmarks (\boyer and
\mqueens), the \gengc collector performs substantially better than the
\incgc collector, but in most cases \gengc uses more memory than
\incgc.

\section{Related work}
\label{sec:relatedwork}

In \cite{PittomvilsBruynoogheWillems@SLP-85}, Pittomvils, Bruynooghe
and Willems present an incremental marking algorithm and show how this
can be used for a generational mark and compact garbage collector.
Their algorithm is capable of marking one heap segment at a time and
can be used in a mark and compact collector to compact one heap
segment during each collection cycle. The generational collector
presented there has a simple policy: data that survived one
collection is promoted to the old generation and the old generation
is never collected. For the collection of one heap segment
independently of other segments, all pointers to heap cells within
that segment are needed.  The trail contains all inter-segment forward
(old to new) pointers, but for inter-segment backward (new to old)
pointers an extra data structure is used: the update table. Each heap
segment is associated with an update table which contains all backward
pointers to that heap segment. The update tables are filled
dynamically during the execution of the program. Since each heap
segment is compacted only once, update tables are only needed for the
most recent segments that have not yet been collected.

The most important difference with our work, is that their approach
uses a heap segment as the basic unit for collections. This has as
advantage that the trail can be used as a remembered set for forward
pointers and separate remembered sets (update tables) are only needed
for backward pointers, while in our approach, all \interblock pointers
are recorded in remembered sets, and some pointers can appear both in
the trail and in a \remset.  However, we believe that using heap
segments as increments is not a good choice for an incremental
collector, since most programs are rather deterministic and the heap
often consists of just a few large segments.
\\

Touati and Hama present their heap layout and collector for the WAM in
\cite{GcLight@FGCS-88}. Their collector is basically a simple
generational collector. The heap is divided into two generations: an
old generation (the old space) and a fixed size nursery space (new
space). The new space is used for new allocations and contains only
data which have never been collected. Garbage collection is triggered
when the new space overflows. During a collection, all live data is
copied from the new space to the old space. The old space is never
collected. The idea behind this collector is to enhance the
performance by using a copying algorithm as much as possible, but
otherwise fall back to a mark and compact algorithm.  This has as
benefits that the order of the heap segments can be preserved and
that, consequently, instant reclaiming can be performed upon
backtracking. During a collection, only the trail, choice point and
environment entries more recent than the last collection are part of
the root set and need to be scanned; only references to heap cells in
new heap space are followed. By keeping the new space small, the
locality of the collector is improved and paging can be reduced.

Our \gengc collector could be configured in a similar way as the
collector by Touati and Hama. However, even with a similar
configuration, there would still be a some important differences.  The
first difference is that their collector does not need remembered
sets. In their system, all variable bindings are recorded on the
trail, so that the trail can be used as a remembered set for
references from the old space to the new space. The second difference
is that their collector only needs to scan the most recent trail,
choice point and environment entries, while our collector needs to scan
all trail, choice point and environment entries at each collection.
Those differences are mainly optimisations which are possible in the
system of Touati and Hama because of the fixed generational policy.
\\

In \cite{Li@PPDP-00}, Li presents chronological garbage collection,
the memory management approach taken in the implementation of the
Logic Virtual Machine (LVM). Just like the WAM \cite{DHWa83,AitK90},
the LVM is an execution model that can be used as a basis for Prolog
implementations. An important difference between the WAM and the LVM
is their memory architecture. Whereas the WAM uses separate stacks for
data structures (the global stack or heap) and choice points and
environments, the LVM merges choice points, environments and data
structures into one global stack.

The chronological garbage collector is a generational copying
collector with particular choices regarding generation organisation,
garbage collection scheduling and survivor promotion. The collector
does not define generations as separate spaces in the heap (stack),
but generations are formed at runtime. Each choice point is a
generation boundary and, additionally, in between choice points, extra
generation boundaries are added if the distance between the top of the
stack and the most recent boundary is smaller than a fixed value, the
cache-limit. The value of the cache-limit depends on the machine cache
size. The space between the top of the stack and the most recent
generation boundary is the nursery space, the rest of the heap is the
generational space.  The trail stack is used as remembered set; it
contains all pointers from old to new that cross generation
boundaries. A garbage collection is scheduled whenever the
continuation frame is part of the generational space and the distance
between the continuation frame or the current choice point (whichever
is more recent) and the top of the stack is bigger than the
cache-limit.

The differences between our work and the work by Li are similar to
the differences between our work and \cite{GcLight@FGCS-88}. In the
system by Li, no separate remembered sets are needed, since the
trail is used as a remembered set and only the most recent trail,
choice point and environment entries need to be scanned at each
collection cycle. There is also a difference regarding garbage
collector policy, the garbage collector from Li rejuvenates data
belonging to the old generation when backtracking brings that data
back to the top of the heap. Our generational collector never
rejuvenates data belonging to the old generation.
\\

In \cite{beltway02}, Blackburn, Jones, McKinley and Moss present the
Beltway framework, a garbage collection framework that generalises
existing copying collectors. In the Beltway framework, the heap
consists of a number of independently collectible regions (called
increments), which are arranged in one or more groups (called belts).
The increments each consist of one or more frames, which are aligned,
contiguous regions of memory. To be able to collect the increments
independently of each other, the garbage collector needs to keep track
of the references between increments. This is done with a write
barrier and remembered sets. The write barrier detects all inter-frame
references, but filters out the unneeded ones (for references between
frames belonging to the same increment). There is one remembered set
for each source-target frame pair.

Looking at the level of frames in the Beltway framework, our approach
is very similar. The fixed size heap blocks, we use, are similar to
frames, both in use and implementation. Also the implementation of a
write barrier and remembered sets is similar. We don't have any higher
level structures like increments and belts however.

\section{Conclusion}
\label{sec:conclusion}

We have presented two incremental copying collectors for WAM-based
Prolog systems: a collector with a purely incremental policy and a
collector with a generational policy. Both collectors are based on the
\newheap, a heap layout different from the standard WAM heap layout in
that the heap consists of a number of equal sized blocks instead of
one contiguous memory area.  This new heap layout requires a number of
modifications to the standard WAM. The empirical evaluation shows
that both systems \incgc and \gengc have good performance. Also, their
collectors result in general in lower pause times and have better
memory usage.


\bibliographystyle{alpha}
\bibliography{bibs}


\end{document}

%% file: tn_table_benchmarks.tex
    \begin{tabular}{lrrrr}
      \toprule
      & \textbf{\textit{heap}}
      & \textbf{\textit{trail}}
      & \textbf{\textit{environment}}
      & \textbf{\textit{choicepoint}}
      \\
      &
      &
      & \textbf{\textit{stack}}
      & \textbf{\textit{stack}}
      \\
      \midrule
      \textit{\browse}
      & 1465139
      & 60004
      & 65059
      & 240057
      \\
      \textit{\boyer}
      & 41397354
      & 24
      & 464
      & 229
      \\
      \textit{\dnamatch}
      & 20937949
      & 24
      & 84
      & 116
      \\
      \textit{\tak}
      & 7507909
      & 24
      & 344
      & 116
      \\
      \textit{\serial}
      & 36750835
      & 24
      & 120
      & 116
      \\
      \textit{\chess}
      & -
      & 1738984
      & 6276977
      & 5214953
      \\
      \textit{\mqueens}
      & -
      & 24
      & 129
      & 116
      \\
      \bottomrule
    \end{tabular}

%% file: tn_table1.tex
\begin{tabular}{lrrrrr}
\toprule
& \oldheaptab & \newheaptab \textbf{\scriptsize 0.5M} & \newheaptab \textbf{\scriptsize 1M} & \newheaptab \textbf{\scriptsize 2M} & \newheaptab \textbf{\scriptsize 4M} \\
\midrule
\textbf{\textit{browsegc}} & 3550 & 3620 & 3574 & 3544 & 3546 \\
\textbf{\textit{boyergc}} & 7258 & 7624 & 7582 & 7584 & 7574 \\
\textbf{\textit{dnamatchgc}} & 1830 & 1862 & 1848 & 1846 & 1844 \\
\textbf{\textit{takgc}} & 1052 & 1096 & 1084 & 1076 & 1068 \\
\textbf{\textit{serialgc}} & 5750 & 6008 & 6104 & 6124 & 6166 \\
\bottomrule
\end{tabular}

%% file: tn_table2.tex
\begin{tabular}{llrrrrr}
\toprule
\multicolumn{2}{c}{}
& \oldheaptab 
& \newheapwbtab \textbf{\scriptsize 0.5M}
& \newheapwbtab \textbf{\scriptsize 1M}
& \newheapwbtab \textbf{\scriptsize 2M}
& \newheapwbtab \textbf{\scriptsize 4M}
\\
\midrule
\textbf{\textit{browsegc}}
 & \textbf{t$_{tot}$}
   & 3550 
   & 4520
   & 4466
   & 3970
   & 3984
\\
 & \textbf{m$_{rs}$}
  &
  & 512704
  & 252556
  & 0
  & 0
\\ \addlinespace[0.2pt]
\textbf{\textit{boyergc}}
 & \textbf{t$_{tot}$}
   & 7258 
   & 7878
   & 7816
   & 7808
   & 7748
\\
 & \textbf{m$_{rs}$}
  &
  & 1114
  & 622
  & 313
  & 177
\\ \addlinespace[0.2pt]
\textbf{\textit{dnamatchgc}}
 & \textbf{t$_{tot}$}
   & 1830 
   & 1928
   & 1920
   & 1918
   & 1908
\\
 & \textbf{m$_{rs}$}
  &
  & 219
  & 108
  & 50
  & 15
\\ \addlinespace[0.2pt]
\textbf{\textit{takgc}}
 & \textbf{t$_{tot}$}
   & 1052 
   & 1090
   & 1084
   & 1086
   & 1074
\\
 & \textbf{m$_{rs}$}
  &
  & 0
  & 0
  & 0
  & 0
\\ \addlinespace[0.2pt]
\textbf{\textit{serialgc}}
 & \textbf{t$_{tot}$}
   & 5750 
   & 6932
   & 6898
   & 6894
   & 6924
\\
 & \textbf{m$_{rs}$}
  &
  & 22891763
  & 22891063
  & 22890703
  & 22395212
\\
\bottomrule
\end{tabular}

%% file: tn_inc_table.tex
\begin{tabular}{lrrrrrr}
\toprule
 & & \optgc & \incgc \textbf{0.5M} & \incgc \textbf{1M} & \incgc \textbf{2M} & \incgc \textbf{4M} \\
\midrule
\textbf{\textit{chess}} \\
   & \textbf{t$_{gc}$/t$_{tot}$}
       & 1877/8528 & 8208/15679 & 3016/10377 & 1329/8661 & 896/8199 \\
   & \textbf{n$_{gc}$}
       & 6 & 51 & 14 & 6 & 3 \\
   & \textbf{t$_{l}$/t$_{a}$/t$_{u}$}
       & 0/312/559 & 26/160/385 & 62/215/345 & 63/221/417 & 149/298/501 \\
   & \textbf{m$_{alloc}$}
       & 8388608 & 10/5242880 & 6/6291456 & 4/8388608 & 3/12582912 \\
   & \textbf{m$_{hp}$}
       & 3997994 & 3897019 & 4186310 & 4633025 & 6342139 \\
   & \textbf{m$_{rs}$}
       & & 641602& 731866& 342425& 571135\\
\cmidrule{1-7}
\textbf{\textit{mqueens}} \\
   & \textbf{t$_{gc}$/t$_{tot}$}
       & 11893/62927 & 25354/75777 & 21673/71761 & 16936/66942 & 10895/60858 \\
   & \textbf{n$_{gc}$}
       & 135 & 1487 & 713 & 326 & 137 \\
   & \textbf{t$_{l}$/t$_{a}$/t$_{u}$}
       & 7/88/268 & 0/17/67 & 2/30/99 & 3/51/172 & 11/79/325 \\
   & \textbf{m$_{alloc}$}
       & 33554432 & 24/12582912 & 12/12582912 & 7/14680064 & 5/20971520 \\
   & \textbf{m$_{hp}$}
       & 15998004 & 12012062 & 11985481 & 12489334 & 14672055 \\
   & \textbf{m$_{rs}$}
       & & 86657& 68445& 35532& 52576\\
\cmidrule{1-7}
\textbf{\textit{boyergc}} \\
   & \textbf{t$_{gc}$/t$_{tot}$}
       & 1374/8730 & 1747/9722 & 1664/9558 & 1226/9029 & 547/8343 \\
   & \textbf{n$_{gc}$}
       & 22 & 147 & 69 & 30 & 10 \\
   & \textbf{t$_{l}$/t$_{a}$/t$_{u}$}
       & 13/62/152 & 2/11/33 & 3/24/66 & 13/40/83 & 15/54/101 \\
   & \textbf{m$_{alloc}$}
       & 16777216 & 13/6815744 & 7/7340032 & 4/8388608 & 3/12582912 \\
   & \textbf{m$_{hp}$}
       & 7997481 & 5482512 & 6069710 & 6469965 & 9794011 \\
   & \textbf{m$_{rs}$}
       & & 80118& 25640& 18688& 115\\
\cmidrule{1-7}
\textbf{\textit{serialgc}} \\
   & \textbf{t$_{gc}$/t$_{tot}$}
       & 1256/6435 & 2983/9724 & 2862/9664 & 2388/8775 & 1622/7996 \\
   & \textbf{n$_{gc}$}
       & 8 & 132 & 66 & 30 & 11 \\
   & \textbf{t$_{l}$/t$_{a}$/t$_{u}$}
       & 48/157/386 & 0/22/116 & 0/43/192 & 1/79/272 & 3/147/550 \\
   & \textbf{m$_{alloc}$}
       & 33554432 & 33/17301504 & 17/17825792 & 7/14680064 & 4/16777216 \\
   & \textbf{m$_{hp}$}
       & 15998002 & 14010666 & 14232282 & 11217807 & 11935801 \\
   & \textbf{m$_{rs}$}
       & & 4432269& 5543526& 4147366& 3955308\\
\bottomrule
\end{tabular}

%% file: tn_gen_table.tex
\begin{tabular}{lrrrrrr}
\toprule
 & & \optgc & \gengc \textbf{0.5M} & \gengc \textbf{1M} & \gengc \textbf{2M} & \gengc \textbf{4M} \\
\midrule
\textbf{\textit{chess}} \\
   & \textbf{t$_{gc}$/t$_{tot}$}
       & 1877/8528 & 5625/13124 & 2998/10386 & 1864/9189 & 1125/8529 \\
   & \textbf{n$_{gc}$}
       & 6 & 36 & 15 & 8 & 4 \\
   & \textbf{t$_{l}$/t$_{a}$/t$_{u}$}
       & 0/312/559 & 0/156/303 & 0/199/362 & 0/233/431 & 0/281/463 \\
   & \textbf{m$_{alloc}$}
       & 8388608 & 12/6291456 & 7/7340032 & 5/10485760 & 4/16777216 \\
   & \textbf{m$_{hp}$}
       & 3997994 & 3783191 & 4946856 & 4320875 & 5696215 \\
   & \textbf{m$_{rs}$}
       & & 846086& 738650& 617326& 207245\\
\cmidrule{1-7}
\textbf{\textit{mqueens}} \\
   & \textbf{t$_{gc}$/t$_{tot}$}
       & 11893/62927 & 1862/51987 & 1764/51765 & 1671/51520 & 1604/51447 \\
   & \textbf{n$_{gc}$}
       & 135 & 791 & 394 & 196 & 98 \\
   & \textbf{t$_{l}$/t$_{a}$/t$_{u}$}
       & 7/88/268 & 1/2/23 & 2/4/47 & 3/8/86 & 6/16/167 \\
   & \textbf{m$_{alloc}$}
       & 33554432 & 21/11010048 & 12/12582912 & 7/14680064 & 5/20971520 \\
   & \textbf{m$_{hp}$}
       & 15998004 & 10073777 & 10576005 & 11594488 & 13645829 \\
   & \textbf{m$_{rs}$}
       & & 65916& 45021& 30418& 21051\\
\cmidrule{1-7}
\textbf{\textit{boyergc}} \\
   & \textbf{t$_{gc}$/t$_{tot}$}
       & 1374/8730 & 1150/9108 & 1096/8953 & 1064/8864 & 658/8431 \\
   & \textbf{n$_{gc}$}
       & 22 & 127 & 62 & 31 & 11 \\
   & \textbf{t$_{l}$/t$_{a}$/t$_{u}$}
       & 13/62/152 & 2/9/20 & 4/17/45 & 10/34/81 & 22/59/110 \\
   & \textbf{m$_{alloc}$}
       & 16777216 & 17/8912896 & 8/8388608 & 5/10485760 & 4/16777216 \\
   & \textbf{m$_{hp}$}
       & 7997481 & 5111256 & 5348241 & 7072758 & 8765287 \\
   & \textbf{m$_{rs}$}
       & & 105562& 33484& 52513& 27310\\
\cmidrule{1-7}
\textbf{\textit{serialgc}} \\
   & \textbf{t$_{gc}$/t$_{tot}$}
       & 1256/6435 & 2495/9946 & 2172/9621 & 2445/9387 & 1835/9000 \\
   & \textbf{n$_{gc}$}
       & 8 & 158 & 80 & 43 & 16 \\
   & \textbf{t$_{l}$/t$_{a}$/t$_{u}$}
       & 48/157/386 & 1/15/46 & 1/27/66 & 3/56/195 & 3/114/505 \\
   & \textbf{m$_{alloc}$}
       & 33554432 & 39/20447232 & 18/18874368 & 10/20971520 & 5/20971520 \\
   & \textbf{m$_{hp}$}
       & 15998002 & 16149064 & 13576387 & 13689080 & 12431808 \\
   & \textbf{m$_{rs}$}
       & & 5866019& 4621922& 4765461& 3621150\\
\bottomrule
\end{tabular}